\documentclass[journal]{IEEEtran}

\usepackage{array}
\usepackage{float}
\usepackage{gensymb}
\usepackage{amsmath}
\usepackage{subfigure}
\usepackage{fixltx2e}
\usepackage{tabularx}
\usepackage[linesnumbered,ruled,vlined]{algorithm2e}
\SetKwComment{Comment}{$\triangleright$\ }{}
\SetKwRepeat{Do}{do}{while}
\usepackage{flushend}
\usepackage{microtype}
\usepackage{graphicx}
\usepackage{caption}
\usepackage{xcolor}

\begin{document}

\title{Spatiotemporal Rate Adaptive Tiled Scheme for 360 Sports Events}

\author{
\IEEEauthorblockN{Tarek El-Ganainy} \\
\IEEEauthorblockA{School of Computing Science \\ Simon Fraser University \\ Burnaby, BC, Canada}
}

\maketitle

\raggedbottom

\begin{abstract}
The recent rise of interest in Virtual Reality (VR) came with the availability of commodity commercial VR products, such as the Head Mounted Displays (HMD) created by Oculus and other vendors. One of the main applications of virtual reality that has been recently adopted is streaming sports events. For instance, the last olympics held in Rio De Janeiro was streamed over the Internet for users to view on VR headsets or using 360 video players \cite{RioVR}. A big challenge for streaming VR sports events is the users limited bandwidth and the amount of data required to transmit 360 videos. While 360 video demands high bandwidth, at any time instant users are only viewing a small portion of the video according to the HMD field of view (FOV). Many approaches have been proposed in the literature such as proposing new representations (e.g. pyramid and offset-cubemap) and tiling the video and streaming the tiles currently being viewed. In this paper, we propose a tiled streaming framework, where we provide a degrading quality model similar to the state-of-the-art offset-cubemap while minimizing its storage requirements at the server side. We conduct objective studies showing the effectiveness of our approach providing smooth degradation of quality from the user FOV to the back of the 360 space. In addition, we conduct subjective studies showing that users tend to prefer our proposed scheme over offset-cubemap in low bandwidth connections, and they don't feel difference for higher bandwidth connections. That is, we achieve better perceived quality with huge storage savings up to 670\%.
\end{abstract}

\begin{IEEEkeywords}
Virtual Reality, Streaming, Projections, Tiling, Multimedia Coding
\end{IEEEkeywords}

\section{Introduction}

Interest in Virtual Reality (VR) is on the rise. Till recently, VR was just being studied at universities labs or research institutes with few failed trials to commercialize it. Many obstacles were in the way of providing end users with VR devices. For instance, the size of the headsets used to be huge, the quality of the screens was low, not much content were specifically designed for VR devices, and the accuracy of head tracking wasn't that good which led to discomfort. Over the years, researchers / industry worked on solving those issues. Nothing major was done till the introduction of Oculus Rift \cite{rift}. Thereafter, many of the major players in the computer industry introduced their own headsets. For example: Google Cardboard \cite{GoogleCB} and Daydream \cite{GoogleDD}, HTC VIVE \cite{HTCvive}, Sony PlayStation VR \cite{SonyVR}, and Samsung GearVR \cite{SamsungGearVR}. With the increasing progress of consumer grade VR headsets a tremendous attention is directed to creating and streaming content to such devices. The headset providers, in addition to many others, are currently producing 360 cameras to enable the creation of VR content. For instance, GoPro Omni \cite{GoProOmni}, Google Odyssey \cite{GoogleOdyssey}, Samsung Project Beyond \cite{SamsungBeyond}, Facebook Surround 360 \cite{FBSurrond}. Similarly, major multimedia streaming service providers such as Facebook \cite{Facebook} and YouTube \cite{YouTube} are currently supporting 360 video streaming for VR devices.

\begin{figure}[H]
    \centering
    \includegraphics[width=.35\textwidth]{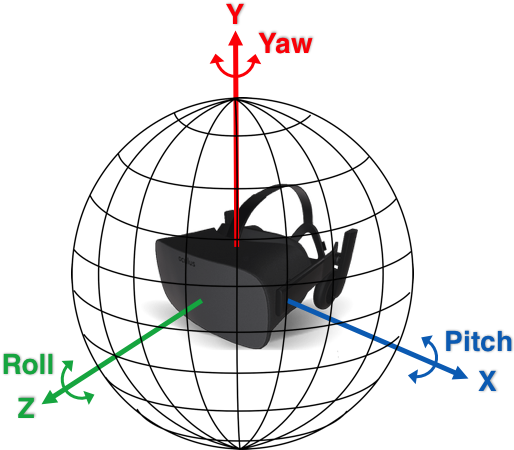}
    \caption{Viewing 360 content on a VR head mounted display.}
    \label{fig:vr_hmd}
\end{figure}

VR content, a.k.a spherical / 360, cannot be viewed with traditional methods. Typically users view VR content on a HMD, such as the Oculus Rift. As shown in Fig. (\ref{fig:vr_hmd}), users can move their heads around the immersive 360 space in all possible directions. Head rotations can be simplified using Euler angles (pitch, yaw, roll) which corresponds to rotations around the (x, y, z) axes respectively. The users viewport can be defined as their head rotation angles, and the field of view (FOV) of the HMD. When the users change their viewport, only the corresponding viewable area of the immersive environment is displayed. That brings us to one of the main challenges in streaming VR content, which is wasting the limited bandwidth on streaming parts of the 360 sphere that are not viewable by the user. This problem exists in other research domains but is crucial in the context of VR, as the content required to deliver immersive experience is very bandwidth demanding (4K+ resolutions, and up to 60 fps). Intuitively, a solution would be to stream only the user's current viewable area. However, if the users move their head fast, which is common in VR, they will experience pauses to buffer the new viewport which strongly affects the immersive experience. Other approaches split the 360 video to tiles, stream the tiles lying in the users FOV with highest bitrate and the rest of tiles with low bitrate. Those approaches suffer from non-uniform assignment of qualities and sharp edges when moving from a high bitrate tile to a low bitrate one. Facebook tried to address these issues by proposing offset-cubemap to offer the user with gradual degradation of quality, where their FOV is projected on the largest number of pixels and gradually decrease the quality (resolution) while turning their head back. Offset-cubemap imposes storage overhead by saving a separate version for a set of predefined user viewports. In this work, we address the problem of assigning degrading qualities to tiles while minimizing the storage requirements.

Our contributions are:
\begin{itemize}
  \item A novel tile segmentation scheme for 360 videos using cubemap projection, namely $tiled\_cubemap$, based on the nature of capturing and viewing immersive videos, specially for sports events.
  \item Formulating the rate adaptation problem for 360 videos utilizing $tiled\_cubemap$.
  \item Proposing a spatiotemporal rate adaptation algorithm for tiled 360 videos.
  \item Objective and subjective experiments to show the merits of our method compared to the state of the art.
\end{itemize}

\section{Background and Related Work} \label{sec:background}

\subsection{VR Content Representation}

VR videos are typically shot using multiple cameras pointing at different directions. The collective field of view for the cameras should cover the 360 space while having enough overlap between them for later stitching purposes. Next, video streams from all the cameras are synchronized, stitched, and projected on a sphere. Finally, to compress the video using standard encoders, we need the video to be in a planar format. Multiple sphere-to-plane mappings have been proposed in to \cite{EquiProj}, \cite{FBCubeMap}, \cite{FBOffsetCubeMap}, \cite{FBPyramid}, \cite{li2016novel}, \cite{fu2009rhombic}. Sphere-to-plane mappings can be categorized to two main categories: (1) Uniform Quality Mappings: all parts of the sphere are mapped to the 2D plane with uniform quality, (2) Variable Quality Mappings: more quality is reserved for areas that users are currently viewing or more likely to view.
\\
\subsubsection{Uniform Quality Mappings}

\begin{figure}[H]
    \centering
    \includegraphics[width=.48\textwidth]{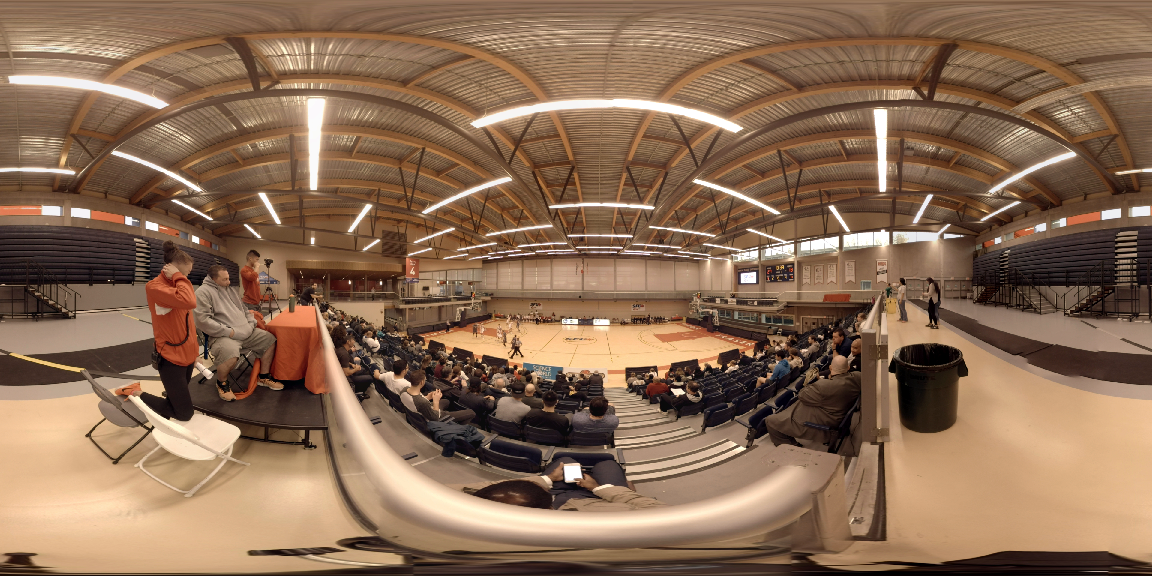}
    \caption{360 video projected as an equirectangular}
    \label{fig:equirectangular}
\end{figure}

\textbf{Equirectangular Projection} is the most common sphere-to-plane mapping. It can be described as unwrapping a sphere on a 2D rectangular plane with the dimensions ($2 \pi r$, $\pi r$), where $r$ is the radius of the sphere. A simple unwrapping of a sphere on a 2D plane will lead to gaps in the output mapping that increase towards the poles. But, for equirectangular projection we stretch the sphere to fit the whole 2D plane as shown in Fig. (\ref{fig:equirectangular}). The most known example for equirectangular projection is the world map. As mentioned above, equirectangular projection is widely supported and easily viewable even with no special players. On the other hand, one of its main drawbacks is the amount of redundant pixels, specially around the poles, which will waste the user's limited bandwidth in a streaming scenario.

\begin{figure}[H]
    \centering
    \includegraphics[width=.48\textwidth]{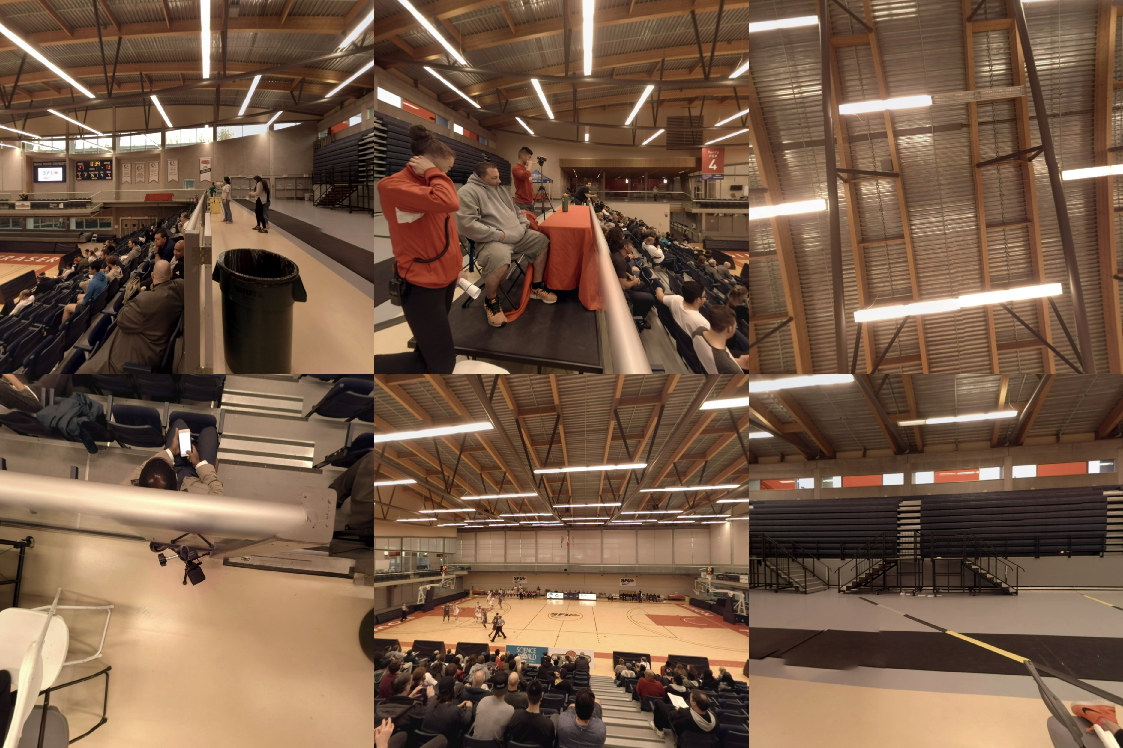}
    \caption{360 video projected as a cubemap. Order of cube faces from top left to bottom right are (right, left, top, bottom, front, back) respectively.}
    \label{fig:cubemap}
\end{figure}

\textbf{Cubemap Projection} has been used extensively in gaming applications, and can be easily rendered using graphics libraries. Similar to equirectangular projection, we are trying to project a sphere on a 2D rectangular plane, but in this case we project the sphere on a cube first. Unlike equirectangular projection, cubemap doesn't have stretched areas. A visual example for cubemap projection is shown in Fig. (\ref{fig:cubemap}). Facebook released an open source implementation for an ffmpeg \cite{FFmpeg} filter to convert a video mapped as an equirectangular to cubemap \cite{FBCubeMapCode}.

\subsubsection{Variable Quality Mappings}

\textbf{Pyramid Projection} is one of the early trials by Facebook to support variable quality mappings \cite{FBPyramid}. The main idea is to project the sphere on a pyramid where its base is the user's current viewing area. By doing so, the user's viewport will be represented with highest number of pixels, and we'll have a degradation of quality as the users move their head to the back. There are two main issues with pyramid projection: (1) as the users rotate their head by $120^{\circ}$, the quality drops the same amount as they turn their head to the back of the pyramid, (2) Since pyramid projection is not supported on GPUs, it's not as efficient to render them as it is for a cubemap.

\begin{figure}[h]
    \centering
    \includegraphics[width=.48\textwidth]{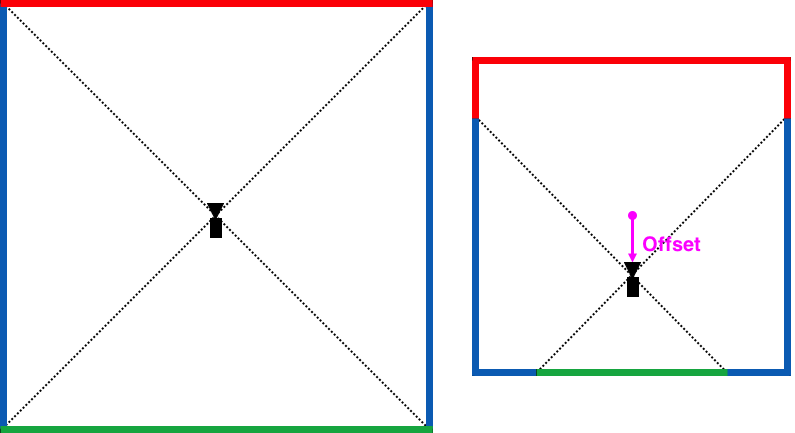}
    \caption{Offset-cubemap illustrated. Cubemap on the left, offset-cubemap on the right. The user FOV is maintained at same quality. In case of offset-cubemap, more pixels are given to the user FOV while the quality degrades gradually when moving to the sides and the back.}
    \label{fig:offset_cubemap_illustrated}
\end{figure}

\begin{figure}[h]
    \centering
    \includegraphics[width=.48\textwidth]{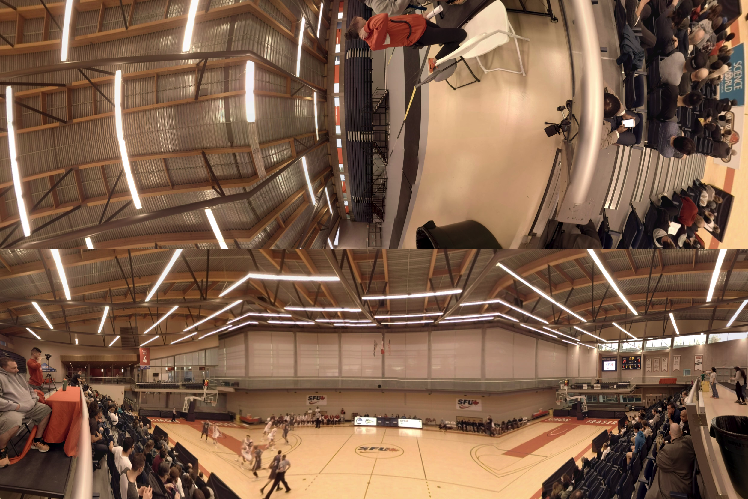}
    \caption{360 video projected as an offset-cubemap in baseball format, where the top part shows the faces (top, back, bottom) while bottom contains the (left, front, right) faces of the cube.}
    \label{fig:offset_cubemap}
\end{figure}

\textbf{Offset-cubemap Projection} is the state-of-the-art variable quality mapping proposed by Facebook, it builds on top of the cubemap to provide a variable quality mapping while addressing the issues of the pyramid projection. Offset-cubemap is a regular cubemap where the user is pushed back from the center of the cube with a predefined offset in the opposite direction from where they are viewing the video as shown in Fig. (\ref{fig:offset_cubemap_illustrated}). An example of a video projected as an offset-cubemap is shown in Fig. (\ref{fig:offset_cubemap}). Since the offset-cubemap is essentially a cubemap, we can efficiently render it on a GPU. Also, offset-cubemaps offer smoother degradation of quality compared to a pyramid projection.  Offset-cubemap isn't widely adopted yet, a recent measurement study \cite{qian2016optimizing} on Facebook's 360 streaming pipeline shows that they are still using the traditional cubemap projection. Recently, Facebook released the implementation of the offset-cubemap as part of the cubemap open source code \cite{FBCubeMapCode}. The main drawback of offset-cubemap is its storage overhead. To stream an offset-cubemap video using common streaming systems, such as MPEG DASH \cite{sodagar2011mpeg}, in addition to having multiple quality representations for the offset-cubemap to serve different bandwidth profiles, we also need to support different user viewports. To handle different viewports, Facebook proposes to create 30 different versions of the same video covering most of the possible user viewports. That is, for each version the the video is projected as an offset-cubemap in the user viewing direction, where the user FOV is having the highest quality. To address varying bandwidth requirements, they propose to encode an independent version with different offset for each network configuration. This process is applied to all the 30 viewports of the same video. In total, we will have to encode and store (30 multiplied by the number of network configurations) versions of the same video.

\subsection{Tiled Streaming}

Other researchers tackle the problem of streaming high-resolution-videos over limited bandwidth by tiling \cite{d2016using}, \cite{zare2016hevc}. That is, the video is partitioned to multiple tiles and we stream the tiles depending on the user's FOV. Tiling has proven effective in domains such as online video lectures \cite{makar2010real}, and sports \cite{gaddam2016tiling}. The common issues in tiling are: criteria for assigning qualities to tiles; support for tiling in existing streaming frameworks; the effect of mixing tiles with different qualities on the perceived quality; the need of multiple decoders at the client side to decode each independent tile; the bitrate overhead due to constrained motion vectors search space.

For tiled streaming, we partition the video into tiles in both the horizontal and vertical directions while assigning uniform resolution levels to them. Later, we stream only the tiles overlapping with the user viewport. In some studies, mixing tiles with different levels of resolutions is considered to provide a full delivery for spherical / panoramic content. The methods employed to choose the tiles resolution levels were rather crude. Moreover, rendering neighboring tiles with different resolutions will lead to visible seams that will affect the perceived quality by users. Yu et al. \cite{yu2015content} focus on the issues mentioned above in their work. First, they partition the equirectangular video to horizontal tiles. Each horizontal tile is assigned a sampling weight based on it's content and users viewing history. Then, based on the bandwidth budget and the sampling weight of each tile, they optimize the bit-allocation for each tile. Secondly, to overcome the seams problem, they add an overlapping margin between each two neighboring tiles. Finally, alpha blending is applied on the overlapping tile margins to reduce the visible seams effect.

D'Acunto et al. \cite{d2016using} make use of the MPEG-DASH Spatial Relationship Description (SRD) \cite{niamut2016mpeg} extensions to support tiled streaming. SRD can describe a video as a spatial collection of synchronized videos. Using SRD they can stream the area users are currently viewing in the highest quality possible based on their available bandwidth. Also, they provide the users with a seamless experience even if they zoom in or pan around. For zoom, the current streamed segment is up-sampled, and displayed to the user while downloading the high quality segment. For pan, they always stream a fallback lowest quality full sized version of the segment, so they can display the corresponding panned area in low quality while downloading the high quality one. Similarly, Le Feuvre et al. \cite{le2016tiled} utilized the MPEG-DASH SRD extensions to perform tiled streaming. They developed an open-source MPEG-DASH tile-based streaming client using the SRD extensions. They experimented with different adaption modes, both independent H.264 tiles and constrained HEVC tiles.

Since our focus in this work is immersive VR multimedia experiences, streaming only a portion of the video using tiled methods won't be sufficient. That is, to provide an immersive experience we need to provide the users with the whole 360 environment around them not to break the sense of presence. To do that while keeping the bandwidth requirement in consideration, we may mix tiles with different resolutions / qualities based on their importance from the user's viewing perspective. Wang et al. \cite{wang2014mixing} studied the effect of mixing tile resolutions on the quality perceived by the users. They conducted a psychophysical study with 50 participants. That is, they show the users tiled videos, where the tile resolutions are chosen randomly from two different levels. The two levels of resolutions are chosen every time so one of them is the original video resolution, and the other is one level lower. The experiments show that in most cases when they mix HD (1920x1080p) tiles with (1600x900p) tiles participants won't notice any difference. Also, when they mix the HD tiles with (960x540p) the majority of the participants are fine with the degradation in quality when viewing low to medium motion videos. Furthermore, for high motion video, more than 40\% of the participants accept the quality degradation.

Zare et al. \cite{zare2016hevc} address the need for having multiple encoders at the client side, and investigate the tiling overhead. They proposed an HEVC-compliant tiled streaming approach utilizing the motion-constrained tile sets (MCTS) concept. MCTS have been used where multiple tiles are merged at the client side to support devices with a single hardware decoder. They propose to have two versions of the video, one in high resolution, and the other in low resolution. The two versions of the video are partitioned to tiles, and streamed to users based on their current viewport. That is, the tiles currently viewed by the user are streamed in high resolution, while the rest of the tiles are streamed in low resolution. While encoding each independent tile, they allow motion vectors to point to other tiles that are usually streamed together. Multiple tiling schemes are examined, while smaller tiles minimize the extra non-viewable data sent, they offer less compression ratio. The results show that they can achieve from 30\% to 40\% bitrate reduction based on the tiling scheme chosen.

\subsection{Discussion}

Streaming 360 content has been addressed in the literature in three ways: (1) streaming all the content in uniform quality mappings; (2) streaming variable quality mappings, like pyramid and offset-cubemap; (3) tiled-streaming using techniques mainly developed for pan-tilt-zoom scenarios. Existing solution suffer from problems such as:
\begin{itemize}
  \item Wasting users bandwidth on streaming full quality videos, while they only view a small portion of the video.
  \item Huge storage and encoding overhead for variable quality mappings, by generating 30 different versions of each video to support possible user viewports. In addition, more storage and encoding overhead in case of generating versions with different offsets for different bandwidth requirements.
  \item Most of the proposed tiled-streaming approaches rely on tiling schemes not tailored for 360 video scenarios, where they don't utilize how the videos were originally captured and the users viewing patterns. Moreover, tiles are treated similarly in terms of encoding regardless of their complexity. This is a big issue for 360 videos, since video complexity varies widely over space and time.
\end{itemize}

\begin{figure*}[t!]
\centering
\subfigure[]{\label{fig:tiled_cubemap_1}\includegraphics[width=.48\textwidth]{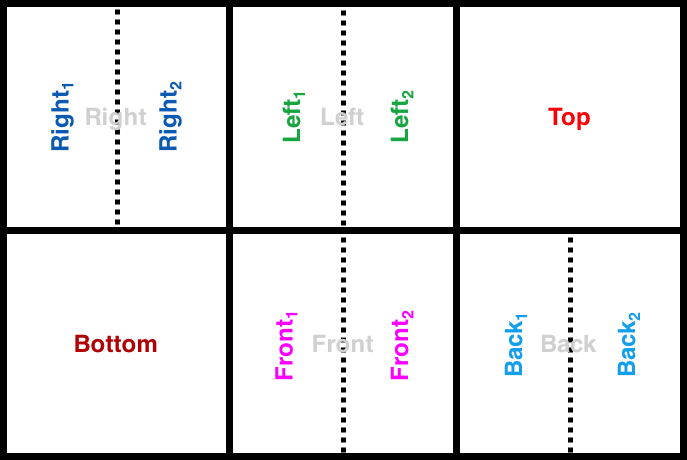}}
\subfigure[]{\label{fig:tiled_cubemap_2}\includegraphics[width=.48\textwidth]{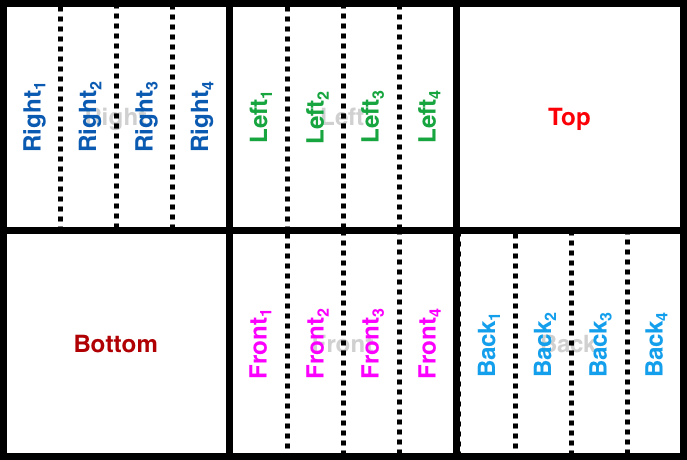}}
\caption{(a) $tiled\_cubemap\textsubscript{1}$, (b) $tiled\_cubemap\textsubscript{2}$} \label{fig:tiled_cubemap}
\end{figure*}

\section{Proposed Framework}
\subsection{Tiled Cubemap}
We propose a novel tiling scheme, namely $tiled\_cubemap$, based on cubemap projection. The intuition behind this tilling scheme is that 360 videos entail high spatial and temporal information. For instance, some parts of the 360 space are less textured, some other parts are highly textured but have less motion, or a mix of both. Also, the top and the bottom faces of the cubemap usually are low in motion/texture, typically sky/ceiling and ground/floor as shown in Fig. (\ref{fig:cubemap}), so we treat them as a single tile. For the rest of the faces we partition them to vertical tiles, where each tile is small enough to offer a smooth gradual degradation of quality while big enough not to impose much encoding overhead from tiling. The intuition behind partitioning tiles to vertical tiles is based on that users vertical FOV in common HMDs is around $90^{\circ}$, which is covered by each tile. For sports events users tend to focus more on the surroundings other then the ceiling and the floor. So, we partition all the tiles to cover $90^{\circ}$ vertical FOV while capturing a typical horizontal head movement $20^{\circ}$ to $45^{\circ}$. We empirically try two variations of the $tiled\_cubemap$, one with 10 tiles ($tiled\_cubemap_1$) and the other with 18 tiles ($tiled\_cubemap_2$) as shown in Fig. (\ref{fig:tiled_cubemap}). Our method relies on optimizing the encoding parameters for each video tile, for each time segment (chunk). We utilize the high spatial nature of 360 videos, as they contain different types of textures and motion patterns by encoding every tile based on its complexity. In addition, we utilize the temporal aspect in 360 videos, as the whole scene may change, or a moving object may change it's place in the 360 space over time. So, we split the video to time chunks and independently decide on the encoding parameter of each chunk based on its complexity.

Our $tiled\_cubemap$ scheme addresses the problems of previous approaches such as traditional tiling scheme or offset-cubemap. First, our tiling scheme is based on cubemap projections, where pixel redundancies are minimal. Additionally, since we are dealing with tiles, there is no need for extra storage for each user viewport, as all the tiles are available in all qualities and they can be mixed and matched on the fly as the client requests a chunk from the server according to its bandwidth requirement and viewport. Moreover, tiled-cubemap is designed while accounting for how the 360 videos are captured and viewed by users, specifically for scenarios such as sports events. Finally, tiled-cubemap encodes each tile based on its complexity, leveraging the spatial properties of 360 videos. That is, quality levels are defined using an objective quality metric, e.g. PSNR. This way the quality level assignment would be more fair compared to fixed quantization parameter (QP) schemes, in which the quality from high textured tiles can be reduced to maintain a fixed bitrate/QP for low texture/static tiles though their perceived quality cannot be increased.
\\
\subsubsection{Per-tile Complexity Analysis}

To demonstrate the need for a temporal per-tile encoding model that we adopt for $tiled\_cubemap$ we perform the following analysis. For simplicity, we use $tiled\_cubemap_1$. We do the analysis on five video sequences for sports events. For each tile we calculate the spatial and temporal information (SI and TI) \cite{itu1999subjective} as following:

\[ SI = median\{\sigma(Sobel(F_n))\} \]
\[ TI = median\{\sigma(M_n(i,j))\} \]

where $F_n$ is frame $n$, and the difference $M_n(i,j)$ between two successive frames $i$ and $j$ is calculated as following:

\[ M_n(i,j) = F_n(i,j) - F_{n-1}(i,j) \]

\begin{figure}[h]
   \centering
   \includegraphics[width=.48\textwidth]{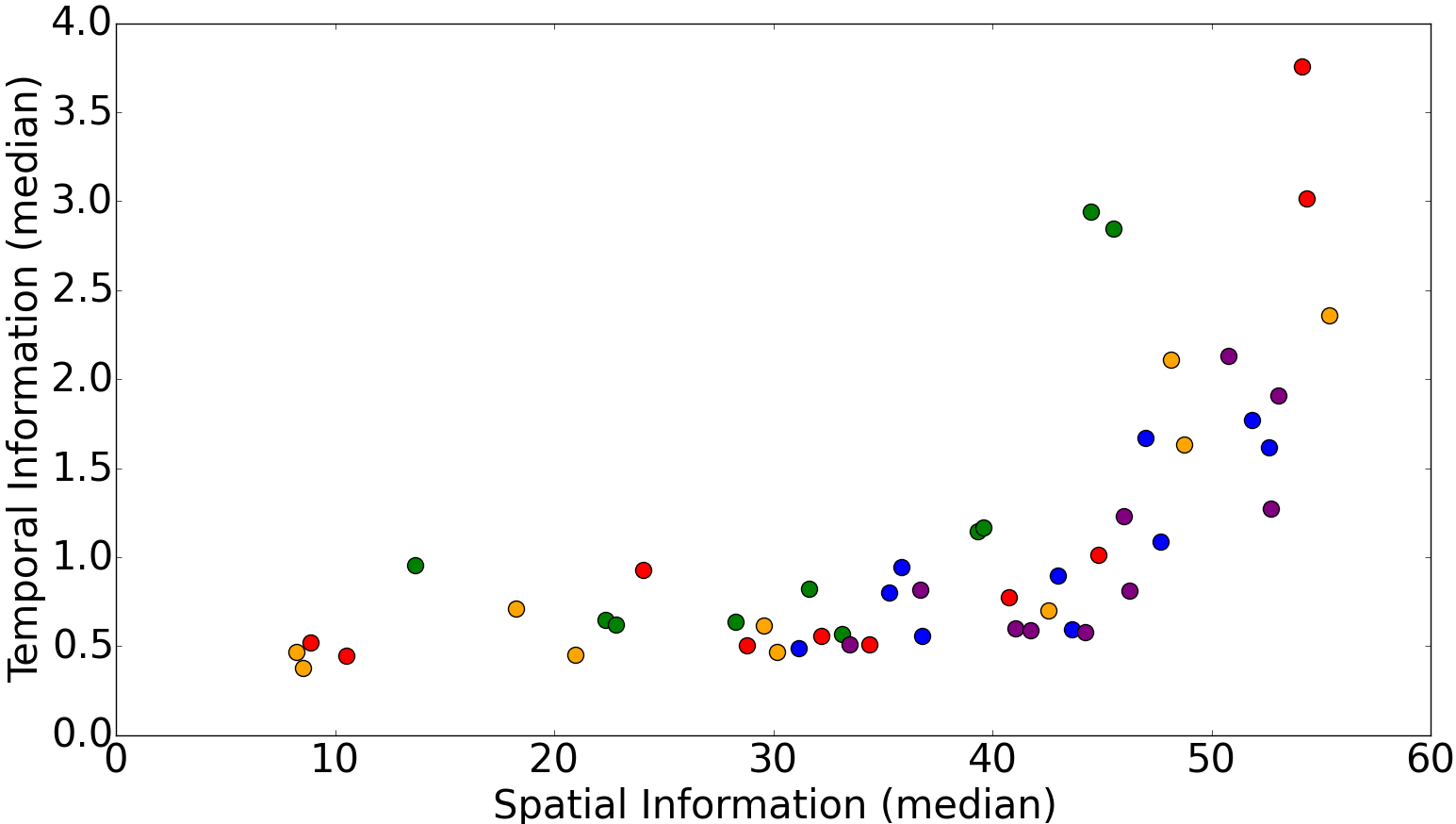}
   \caption{SI, and TI values for all the tiles of 5 different videos (each denoted by a different color) using $tiled\_cubemap_1$. Results indicate high variance in tiles complexity due to the high spatial/temporal nature of the 360 videos.}
   \label{fig:si_ti}
\end{figure}

\begin{figure}[h]
   \centering
   \includegraphics[width=.48\textwidth]{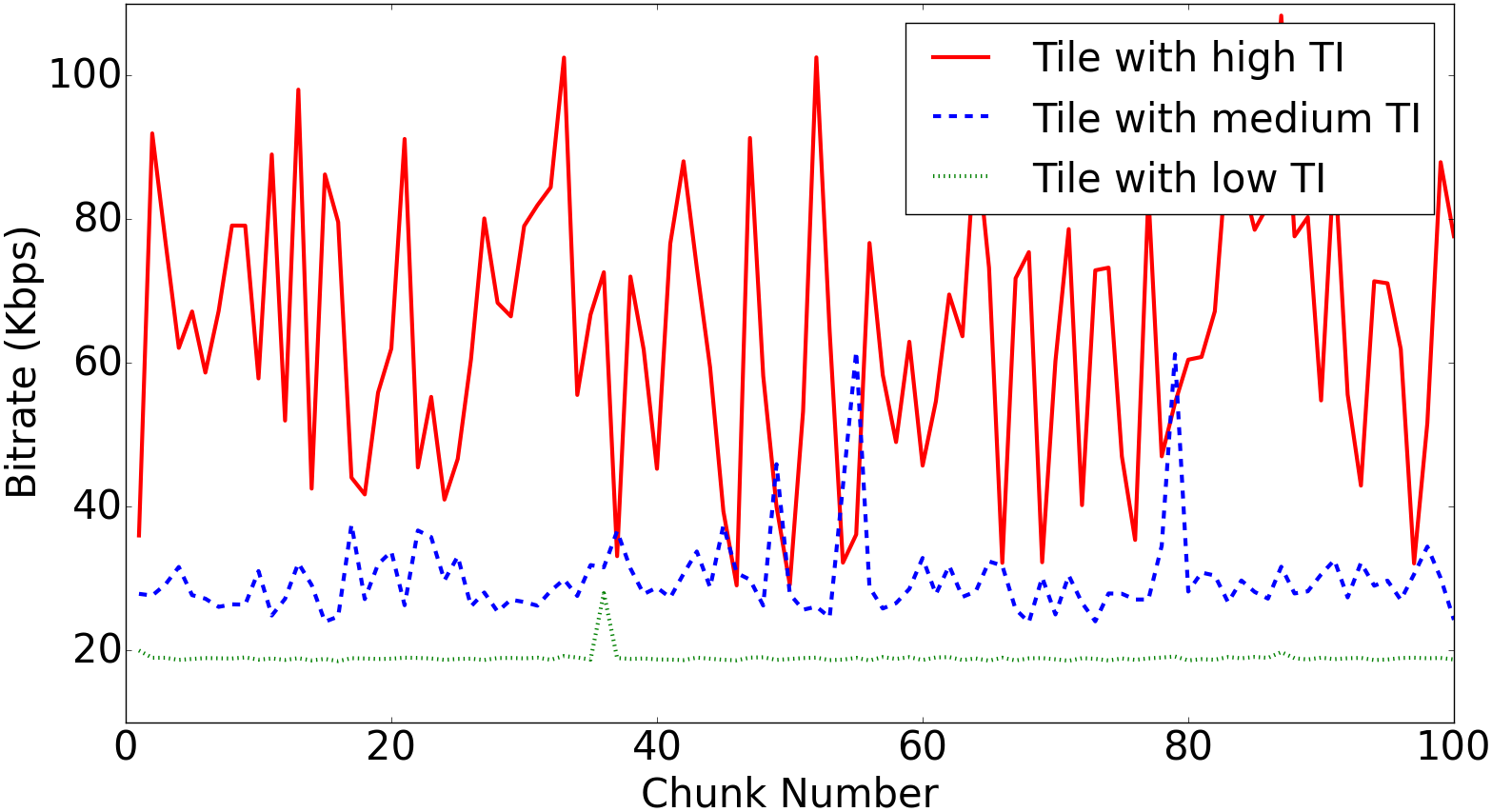}
   \caption{For a fixed PSNR value = 40 dB, frequency of bitrate variations is strongly correlated to temporal information of the tiles.}
   \label{fig:bitrate_overtime}
\end{figure}

As shown in Fig. (\ref{fig:si_ti}) the SI and TI values of all tiles for all videos are widely spread, indicating different spatial and temporal information for each separate tile.

Different spatial and temporal complexities affect the encoding performance of each tile. For instance, we chose three tiles with different TI values (high, medium low) and calculate the bitrate for each tile overtime while fixing the desired PSNR value to 40 dB. As shown in Fig. (\ref{fig:bitrate_overtime}) the bitrate value changes relative to the TI value.

\begin{table}[H]
\begin{center}
\begin{tabular}{lcccccc}
\firsthline
\multicolumn{1}{l}{} & \multicolumn{2}{c}{$\mathbf{QP = 24}$} & \multicolumn{2}{c}{$\mathbf{QP = 36}$}  & \multicolumn{2}{c}{$\mathbf{QP = 45}$}\\
\cline{2-7}
$\mathbf{Tile}$ & $\mathbf{BR\%}$ & $\mathbf{PSNR}$ & $\mathbf{BR\%}$ & $\mathbf{PSNR}$ & $\mathbf{BR\%}$ & $\mathbf{PSNR}$ \\
\hline
$right_1$ & 57.0 & 46.74 & 82.1 & 40.47 & 84.9 & 35.17\\
$right_2$ & 58.2 & 50.24 & 81.1 & 44.91 & 82.1 & 39.31\\
$left_1$ & 57.0 & 48.58 & 82.8 & 42.58 & 85.5 & 37.27\\
$left_2$ & 52.0 & 45.72 & 81.3 & 39.20 & 87.3 & 33.86\\
$top$ & 62.0 & 47.31 & 86.7 & 40.91 & 89.5 & 35.41\\
$bottom$ & 58.5 & 50.02 & 88.1 & 44.61 & 90.8 & 39.91\\
$\boldsymbol{front_1}$ & 49.3 & \textbf{45.43} & 81.6 & \textbf{38.51} & 88.6 & \textbf{33.20}\\
$\boldsymbol{front_2}$ & 51.4 & \textbf{45.70} & 82.8 & \textbf{39.01} & 89.0 & \textbf{33.73}\\
$back_1$ & 58.3 & 51.61 & 78.6 & 46.66 & 79.1 & 40.04\\
$back_2$ & 57.3 & 51.62 & 80.8 & 46.73 & 80.9 & 39.69\\

\lasthline
\end{tabular}
\caption {PSNR and bitrate reduction values when using three QP values to encode different tiles of the same video.} \label{tab:bitrate_PSNR}
\end{center}
\end {table}

Similarly, we encoded all the tiles with different QP values and measure the resulting PSNR and bitrate reduction. Table. (\ref{tab:bitrate_PSNR}) shows how each tile responds differently in terms of bitrate reduction and PSNR based on its spatial location in the 360 space. Encoding all the tiles using the same QP value may result in more bitrate reduction for some tiles compared to others while maintaining higher PSNR value. For instance, as shown in Table. (\ref{tab:bitrate_PSNR}) we can achieve a PSNR value of 45 dB for tile $right_2$ with a QP value of 36 and 81\% bitrate reduction. On the other hand, for tile $front_1$ to achieve the same PSNR value, we have to use QP value of 24, resulting in less bitrate reduction (49\%).

\subsection{Spatiotemporal Rate Adaptation Algorithm}

Now that we have a tiling scheme that accounts for how the video is captured and likely viewed, next is how to allocate the user bandwidth to the tiles. Optimally, we want to stream all the tiles with the highest quality. However, due to the bandwidth constraint that will not be possible. To address this issue, we propose a novel spatiotemporal rate adaptation algorithm targeting $tiled\_cubemap$. That is, our goal is to maximize the overall quality of the video streamed under limited bandwidth, while the user FOV is streamed at the highest possible quality ($Q_{max}$) with a gradual degradation of quality for the rest of the $tiled\_cubemap$. In our algorithm we assume the gradual degradation of quality to follow a normal distribution with steepness ($\sigma$). We choose a normal distribution to offer a smooth gradual degradation of quality. The rate adaptation method is modeled as follows: First, we define a set of quality levels, where $Q(k)$ is the quality of tile $k$. The quality levels are represented as integer values starting from $0$ with an increment of one, where $0$ is the lowest quality. By assigning generic quality levels we have the flexibility to use any video quality metric with our method. Second, we assign priorities to the tiles based on their viewing likelihood for the current user viewport, where $P(k)$ is the priority of tile $k$. Priorities are integer values starting from $0$ with increment of one, where $0$ is the highest priority. For each tile the priority can be assigned in multiple ways, which gives our method the flexibility to adapt to different priority models. In this work, we assume priority levels assigned in a gradual degradation fashion starting with the FOV tiles having the highest priority, and gradually decreasing the priority as we move back. The top and the bottom tiles are assigned priorities similar to the next neighbouring tiles to the user FOV. Each tile quality will depend on the available bandwidth and the priority assigned to it. Depending on the current viewport, the tiles overlapping with the user's FOV have the highest priority ($P(k)=0$), we increment the value of $P(k)$ by 1 as we move to the next set of neighbouring tiles. Finally, we try to maximize the collective qualities of all the tiles, weighted by each tile area $A(k)$, while accounting for their priorities and the bandwidth constraints. We formulate the rate adaptation method as a maximization problem as shown in Eq. (\ref{eq:bitrate_optimization}):

\begin{equation} \label{eq:bitrate_optimization}
\begin{aligned}
& \underset{\sigma}{\arg\max}
& & \Sigma_K{A(k)\,Q_k(\sigma)} \\
& \text{subject to}
&& Q_k(\sigma) = Q_{max}\,e^{-\frac{P(k)^2}{2\sigma^2}}, \\
&&& \Sigma_K{r[tile_k, Q_k(\sigma)]} \leq B, \\
&&& 0 \leq Q_k(\sigma) \leq Q_{max}, \\
&&& \sigma > 0.
\end{aligned}
\end{equation}

Where $r[tile_k, Q_k(\sigma)]$ is the bitrate of tile $k$ with quality level $Q_k(\sigma)$. The 360 video is split into time chunks ($C$). We run the bitrate optimization for each chunk $c_t$, while assigning the tiles priorities based on the user viewport. The optimizer assigns the highest quality to the FOV tiles, then tries to increase the steepness of the quality degradation curve to account for higher qualities for the rest of the tiles as much as the bandwidth allows. We face three problems with the formulation in Eq. (\ref{eq:bitrate_optimization}). First, the optimization problem is non-linear. Second, the qualities are discrete values. Last, if the bandwidth won't allow the FOV tiles to be streamed with $Q_{max}$ even with all the other tiles being at the lowest quality, we need to adjust $Q_{max}$ manually and run the optimization again. To address the issues above, we design a heuristic algorithm shown in Alg. (\ref{alg:optimize_cubemap}) based on the optimization problem.

\begin{algorithm}
  \caption{Rate Adaptation Algorithm for $tiled\_cubemap$}
  \label{alg:optimize_cubemap}
  \SetKwProg{RateAdaption}{RateAdaption}{}{}

  \RateAdaption{$(C, B, \sigma_{step} , Q_{max})$}{
    \ForEach{$c_t \in C$}{%
      Init()\Comment*[r]{$\sigma \gets 0.1; \sigma_{max} \gets 0;$}
      \Do {$U < \Sigma_K{A(k)\,Q_{max}}$}{
        $U \gets \Sigma_K{A(k)\,Q_k(\sigma)}$\Comment*[r]{$Q_k(\sigma) \gets Q_{max}\,e^{-\frac{P(k)^2}{2\sigma^2}}$}
        \uIf{$\Sigma_K{r[tile_k, Q_k(\sigma)]} \leq B$}{
          $\sigma_{max}\gets\sigma$\;
          $\sigma\gets\sigma+\sigma_{step}$\;
        }
        \Else{
          \uIf{$\sigma_{max}=0$ and $Q_{max}>0$}{
            Init()\;
            $Q_{max} \gets Q_{max} - 1$\;
          }
          \Else{
            \textbf{break}\Comment*[r]{Exceeded bandwidth}
          }
        }
      }
      \ForEach{$t_k \in c_t$}{%
        $Q[c_t,t_k] \gets Q_k(\sigma_{max})$\;
      }
    }
    
    \KwRet{$Q[C,T]$}\;
  }
\end{algorithm}

As shown in Alg. (\ref{alg:optimize_cubemap}), the aim is to maximize the user experience ($U$). Initially, for each video chunk we assign priorities to tiles based on the user's viewport. Also, we set the highest steepness to the quality degradation curve using the lowest $\sigma$ value. That in turn result in assigning highest quality $Q_{max}$ to the tiles with highest priority and minimal quality to the rest of the tiles. In case the bandwidth requirements are not met, we decrement $Q_{max}$ one quality level and re-examine the bandwidth requirement. Once we have the $Q_{max}$ set to maximum possible value, we gradually decrease the steepness of the quality curve to maximize the overall quality of all the tiles streamed to the user. To address discrete quality levels, we assume the rounded value of all the $Q(k)$ values to the nearest integer. The algorithm terminates when we cannot increase the tiles qualities without violating the bandwidth limit.

\section{Evaluation}

\subsection{Experimental Setup}

The two main model parameters we had to choose in our experiments are the quality levels to represent the tiles, and the bandwidth profiles. For the quality levels, we decided to use PSNR levels for their wide use and familiarity for readers. Saying that, our model can use any other video quality metric (e.g. SSIM, VQM, etc..). We represent the quality levels in our algorithm as integer values from (0 to 5), where 5 is the highest quality level. We choose the PSNR values corresponding to the quality levels to be (48, 45, 42, 40, 39, 38) dB. For the bandwidth profiles, we checked Akamai's state of the internet report for Q4 2016 \cite{akamai}. According to the report, the global average connection speed is 7 Mbps. They also mention that the global adoption for broadband speeds 4 Mbps, 10 Mbps, and 15 Mbps to be 79\%, 42\%, and 25\% respectively. Note that average broadband speed in areas like the middle-east and Africa can go down to 2 Mbps. In our work, we use three bandwidth profiles: 2 Mbps ($BW_{low}$), 4 Mbps ($BW_{medium}$), and 10 Mbps ($BW_{high}$).

In this work, we used a 360 video dataset for sports events that we captured ourselves using GoPro Omni camera rig. In total, we have five videos for three different sports (Basketball, Hockey, Volleyball), each video is about 15 minutes. The videos resolution is 4K, frame rate 30 fps, and bitrate 9 Mbps on average. We first split each video temporally to 4 seconds chunks, then for each chunk we split it spatially to tiles. Next, we encode the tiles using different QP values (from 18 to 51 with a step of 3). For each tile, we draw its RD curve and choose the versions that corresponds to the PSNR quality levels we use.

We split the videos temporally and spatially and encode them using ffmpeg. For the offset-cubemap implementation, we use the open source code offered by Facebook \cite{FBOffsetCubeMap}. The rate adaptation algorithm is implemented in Python. For the subjective study, we use Oculus Rift HMD to display the videos to the participants. To play the cubemap, and offset-cubemap videos on Oculus Rift, we implement a 360 video player using Unity 3D \cite{unity} that supports both formats.

\subsection{Tiled Cubemap Performance}

\begin{figure}[h]
   \centering
   \includegraphics[width=.48\textwidth]{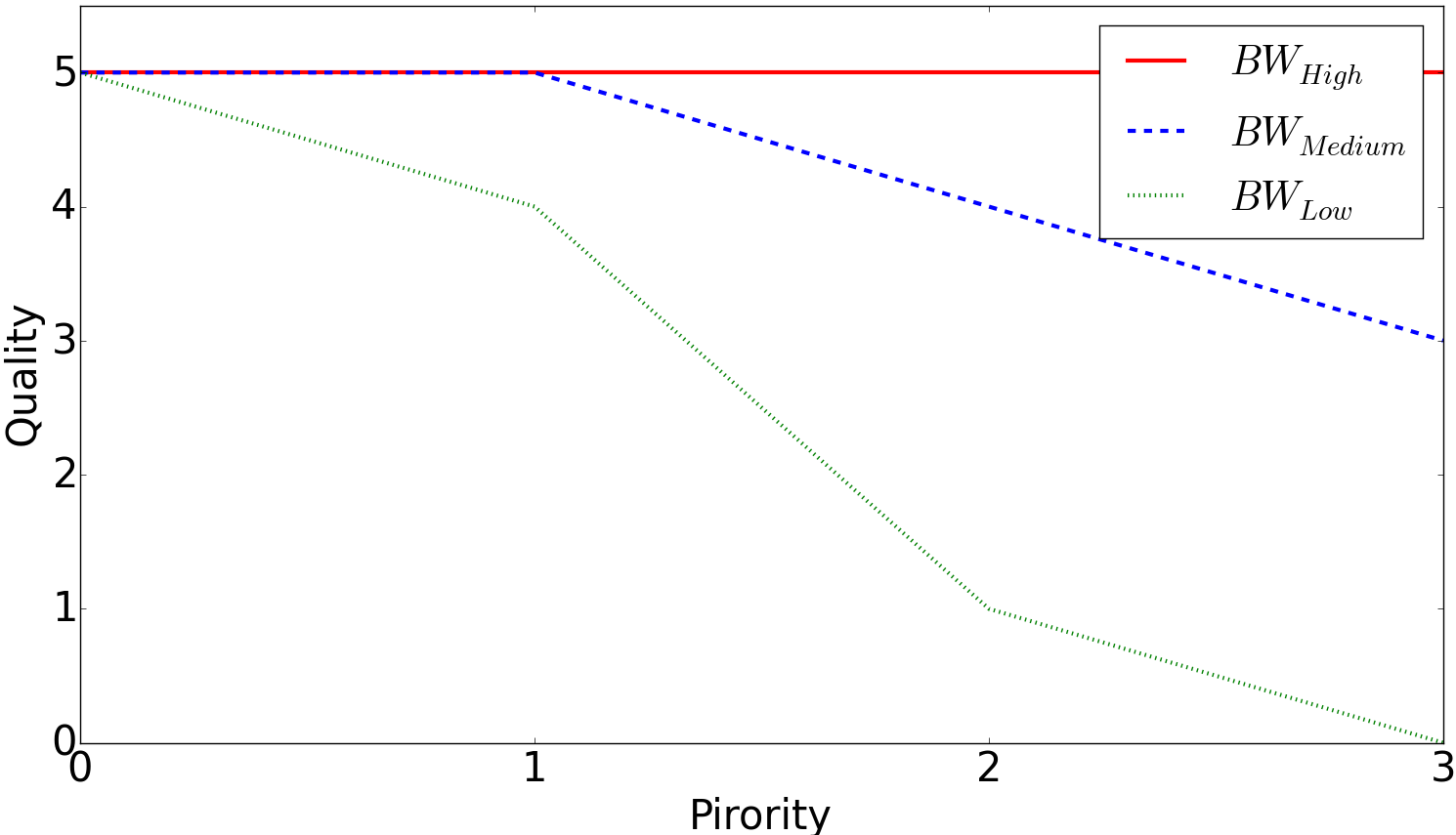}
   \caption{Average quality values (for all test videos, and all time chunks) assigned to different tiles (using $tiled\_cubemap_1$) based on their priority under different bandwidth configurations.}
   \label{fig:perf_tiled_cubemap_1}
\end{figure}

\begin{figure}[h]
   \centering
   \includegraphics[width=.48\textwidth]{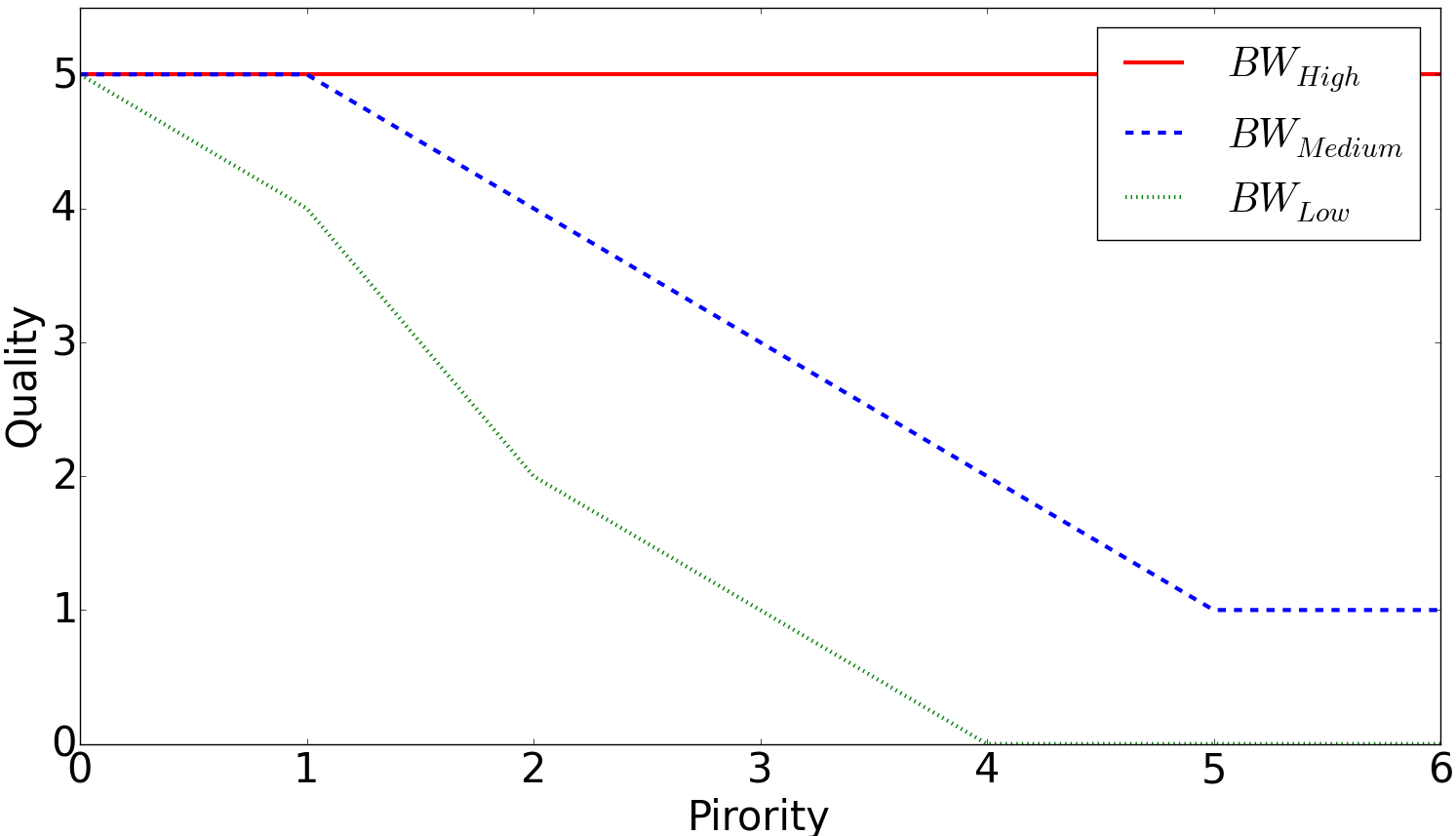}
   \caption{Average quality values (for all test videos, and all time chunks) assigned to different tiles (using $tiled\_cubemap_2$) based on their priority under different bandwidth configurations.}
   \label{fig:perf_tiled_cubemap_2}
\end{figure}

To demonstrate the effectiveness of our method, we run the rate adaptation algorithm for $tiled\_cubemap_{(1,2)}$ under three different network conditions (low, medium, high). In Fig. (\ref{fig:perf_tiled_cubemap_1}), and Fig. (\ref{fig:perf_tiled_cubemap_2}) we show how the rate adaptation algorithm assigns qualities to tiles based on their priorities while meeting the bandwidth requirement. It can be seen how the highest quality is assigned to the tiles with top priority (FOV tiles), while the quality gradually degrades as we move to the less priority tiles. Moreover, we can see as the bandwidth increases the steepness of the quality/priority curve decreases giving more qualities to tiles with less priority. When the bandwidth is large enough, all tiles are chosen with the highest quality.

\begin{figure}[h]
   \centering
   \includegraphics[width=.48\textwidth]{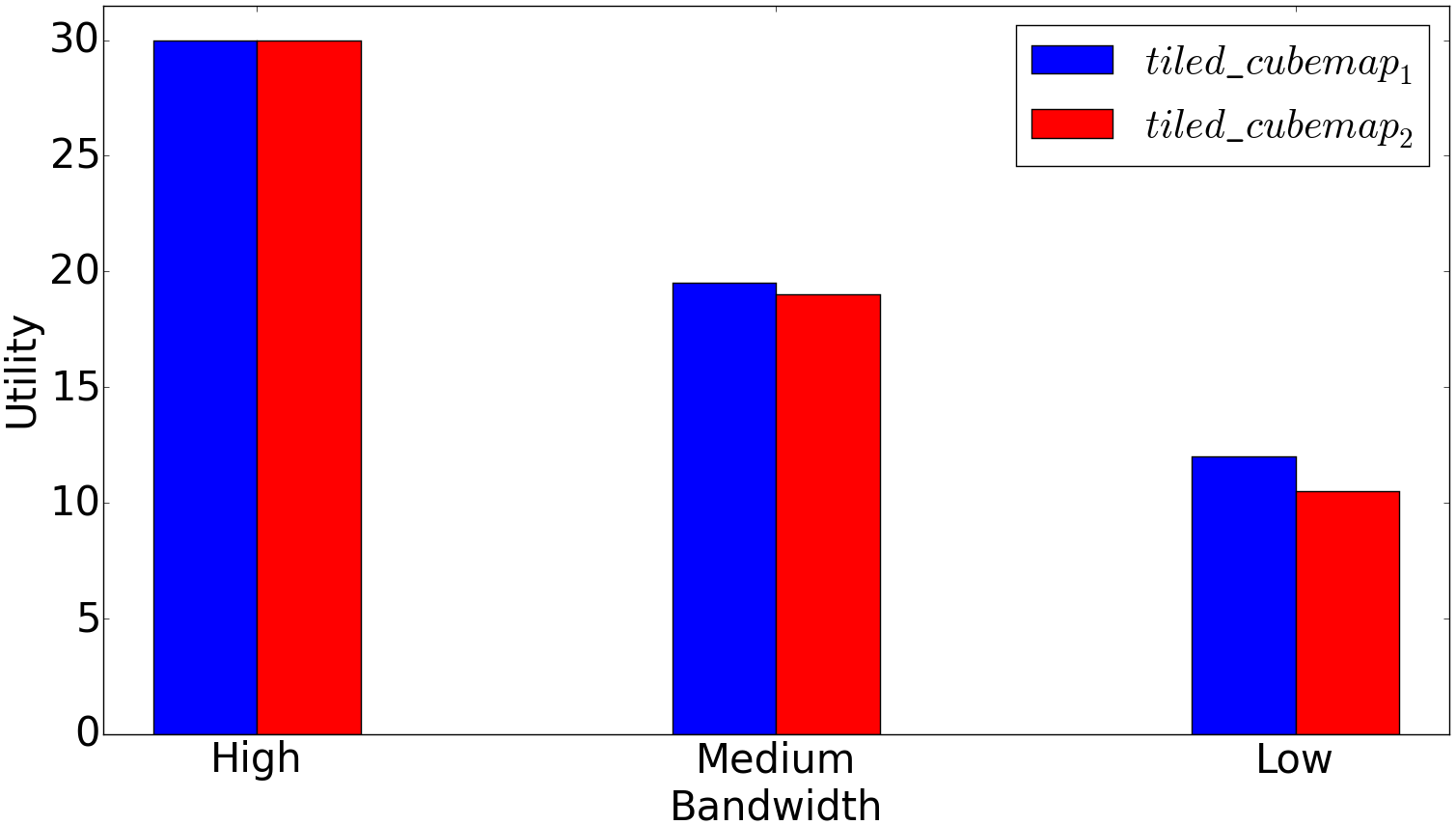}
   \caption{Comparison of the utility values when using $tiled\_cubemap_1$ and $tiled\_cubemap_2$ under different bandwidth configurations.}
   \label{fig:perf_tiled_cubemap_compare}
\end{figure}

One observation is that quality curve is steeper for $tiled\_cubemap_2$ than it is for $tiled\_cubemap_1$. By calculating the utility of the adaptation algorithm using $tiled\_cubemap_1$ and $tiled\_cubemap_2$ as shown in Fig. (\ref{fig:perf_tiled_cubemap_compare}), we see that the utility for $tiled\_cubemap_1$ is higher compared to $tiled\_cubemap_2$. The reduction in quality is a result of the bitrate overhead imposed by tiling. We analyzed the bitrate overhead for both schemes, and found that $tiled\_cubemap_1$ impose $1\%$ bitrate overhead compared to $6\%$ for $tiled\_cubemap_2$. For the rest of the evaluation we will use $tiled\_cubemap_1$ and reference it as $tiled\_cubemap$.

\subsection{Quality Representation}

\begin{figure}[h]
   \centering
   \includegraphics[width=.48\textwidth]{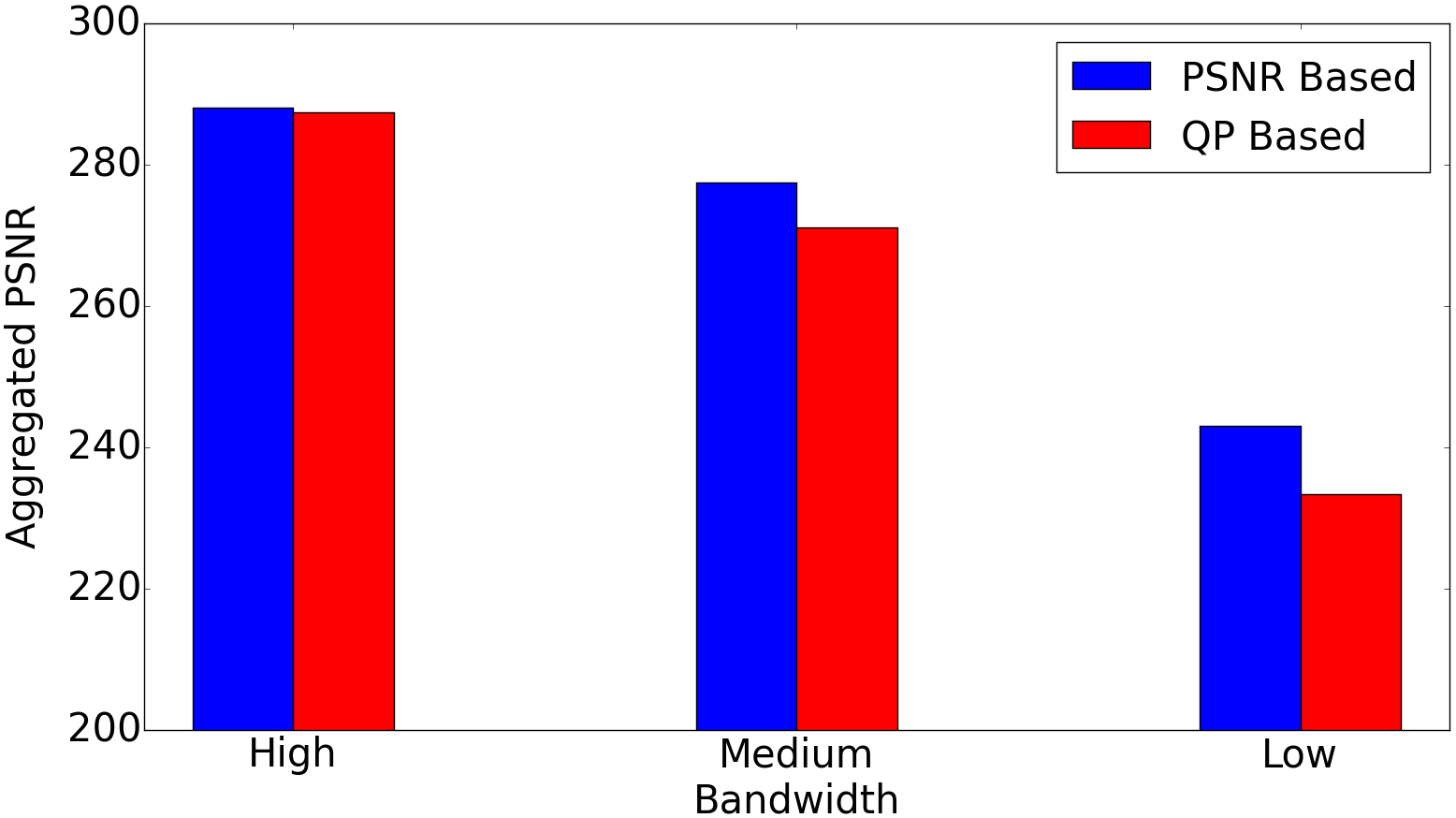}
   \caption{Comparison of the aggregated PSNR values for the six faces of the cube when using a quality based (PSNR) rate adaptation against a QP based one.}
   \label{fig:perf_tiled_cubemap_qp_psnr}
\end{figure}

To evaluate our quality based rate adaptation algorithm against a traditional QP based approach, we conduct a set of experiments under different network profiles and compare the aggregated PSNR of all the tiles. For the QP based approach the quality levels correspond to QP values. In this set of experiments we set the QP values to be (23, 27, 33, 36, 40, 44). The QP values are chosen based on the average QP value used to produce the corresponding PSNR quality level in the PSNR based approach. We run the rate adaptation algorithm for all the videos in our dataset and calculate the average of the aggregated PSNR value for the six faces of the cube. As shown in Fig. (\ref{fig:perf_tiled_cubemap_qp_psnr}), our PSNR based approach achieves consistently better quality than a QP based approach. For instance, in a low bandwidth configuration, we achieve up 1.5 dB improvement per each face of the cube.

\subsection{Comparison Against Offset-cubemap}

\begin{figure}[h]
   \centering
   \includegraphics[width=.48\textwidth]{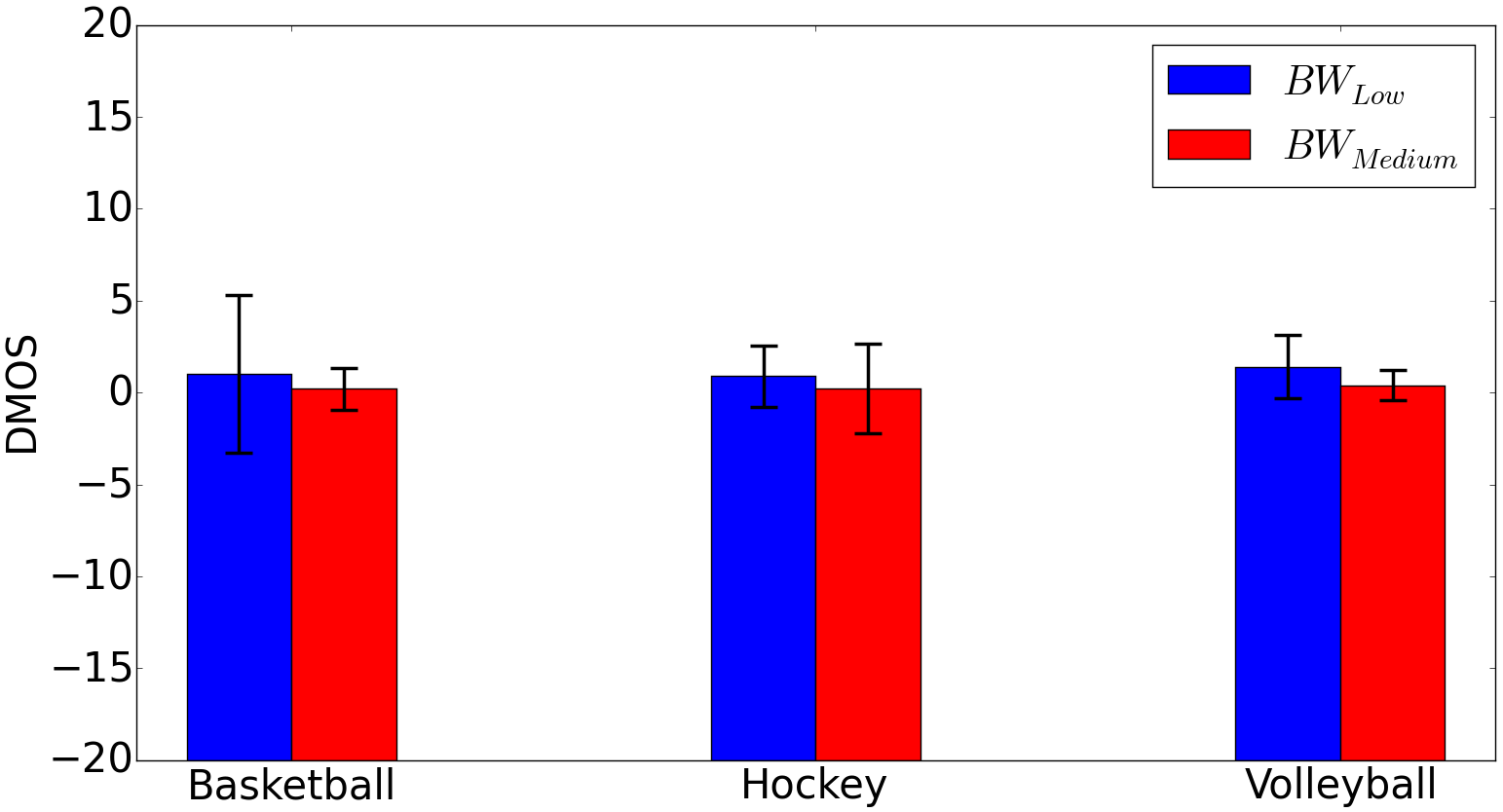}
   \caption{DMOS between the quality of our $tiled\_cubemap$ rate adaptation method and the offset-cubemap.}
   \label{fig:subjective}
\end{figure}

As discussed in Sec. (\ref{sec:background}) offset-cubemap is the most similar representation to our proposed $tiled\_cubemap$. Offset-cubemap offers the users with the highest possible quality in their FOV area and gradual degradation of quality, while adjusting to bandwidth changes by changing the offset value. That is, the quality of areas outside the user FOV will degrade more rapidly as we increase the offset. Similarly, in our proposed $tiled\_cubemap$ scheme, the rate adaptation algorithm assigns the best possible quality to the user FOV, and changes the steepness of the gradual degradation in quality for the rest of the tiles based on the user's available bandwidth. Since offset-cubemap is a non-uniform representation for the video, it is not straight forward to compare it against a uniform representation scheme with an objective measure. We compare our proposed $tiled\_cubemap$ against offset-cubemap by conducting subjective tests. We choose 20 seconds segments from three of our test videos for different sports events (Basketball, Hockey, Volleyball). For each segment, we run our rate adaptation algorithm to choose the proper quality for all tiles based on the users viewport. Likewise, we encode the same segment as an offset-cubemap using different offsets to fit the given bandwidth based on the user viewport. We had 10 graduate students participate in our subjective study. Each participant was asked to view two versions of the same video, one using our method, and the other using offset-cubemap. The goal of the study is to assess the quality of each of the processed videos. All the participants were informed to move their head around to have the highest coverage of the 360 space to assess the differences between the two methods fairly. In this experiment, we apply the double stimulus method (DSCQS), where we show the participants the two processed videos in random order as many times as they need to make an assessment. Then, they were asked to rate the quality of both videos using the standard ITU continuous scale. Finally, we calculate the mean of difference opinion scores (DMOS) by averaging the differences between the scores they gave for our method and those for offset-cubemap. A positive DMOS value indicates that they preferred our method over offset-cubemap, and a negative value indicates otherwise. Results of the subjective study are presented in Fig. (\ref{fig:subjective}). From the results we can see that the participants slightly prefer our $tiled\_cubemap$ when the bandwidth is low, while most participants could not see much difference when a better bandwidth was set. Note that we didn't conduct the experiment for $BW_{high}$ since the video can be streamed at highest quality for both methods.

\begin{figure}[h]
   \centering
   \includegraphics[width=.48\textwidth]{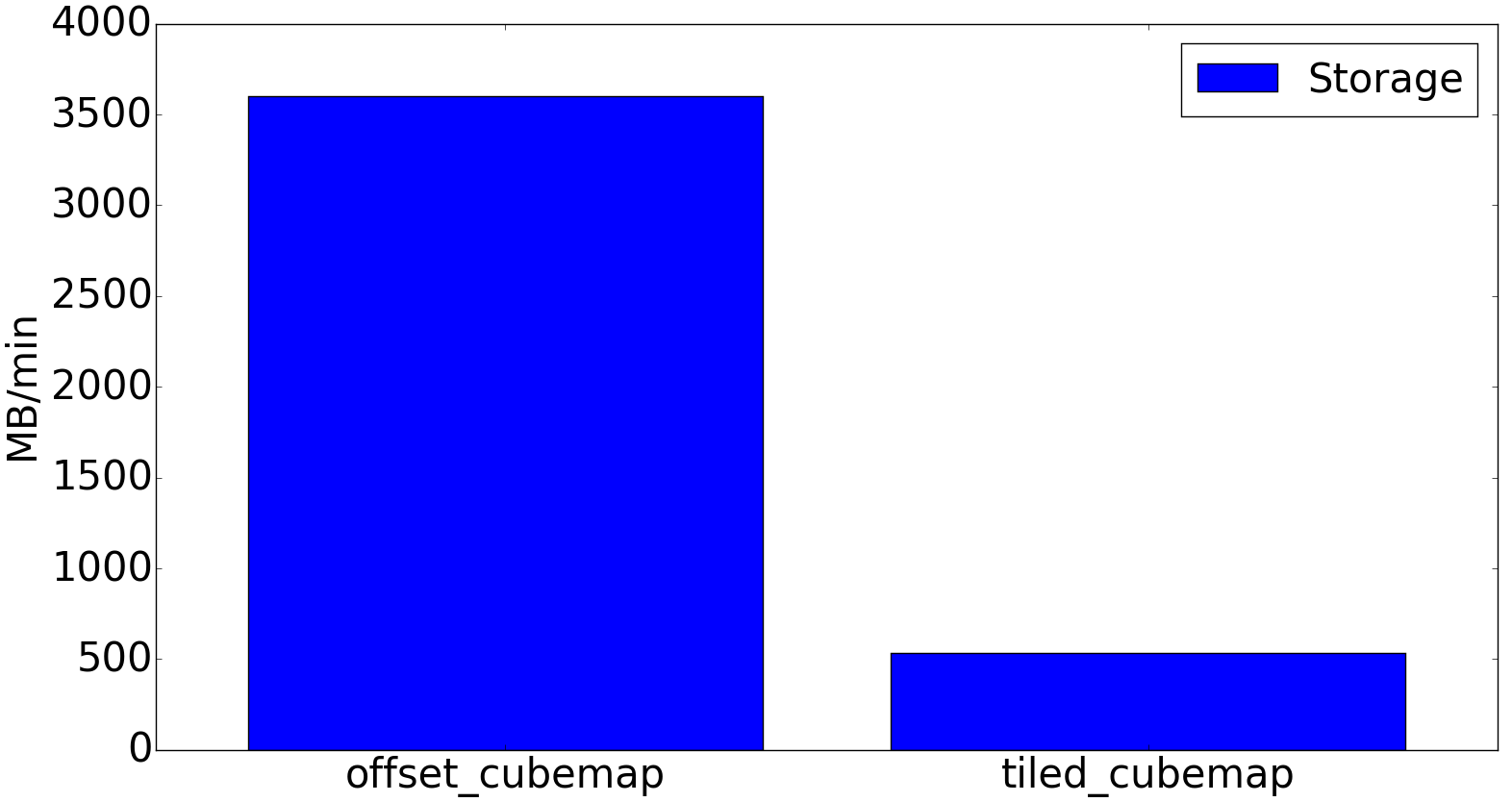}
   \caption{Storage requirements for offset-cubemap and $tiled\_cubemap$ measured in MB/min assuming three bandwidth profiles.}
   \label{fig:storage}
\end{figure}

As shown above, our method provides a perceived quality similar to offset-cubemap. In addition, we offer the flexibility of assigning priorities to tiles based on their content or viewing history. Moreover, the $tiled\_cubemap$ scheme significantly reduce the storage on the server. For offset-cubemap each viewport is represented as an independently encoded version of the same video and saved on disk. As per Facebook, they require 30 different viewports to cover the users potential head movements. Also, if we assume three bandwidth profiles, as in our study, for each viewport we need to process and store three versions with three different offsets. That is, to encode a 360 video as offset-cubemap, the server has to store 90 different versions of the same video (in case of three bandwidth profiles). On the other hand, for $tiled\_cubemap$ after cutting the video to tiles, each tile is stored only six times (the number of quality levels). In Fig. (\ref{fig:storage}) we show the storage requirement per minute for both offset-cubemap and $tiled\_cubemap$ using our video dataset and for the three bandwidth profiles chosen. As seen in the figure, using $tiled\_cubemap$ significantly reduces the storage requirements on the server (up to 670\%).

\section{Conclusion}

In this paper we proposed a novel tiling scheme ($tiled\_cubemap$), targeted mainly for 360 sports events. In addition, based on analyzing our 360 video dataset, we propose utilizing the rich spatial and temporal aspect of 360 videos to make a more efficient use of bandwidth. More specifically, we employ a quality based model for quality levels rather than a typical QP model. We also present a rate adaptation algorithm tailored for $tiled\_cubemap$ to choose the quality level for each tile given the user bandwidth. Finally, we conduct objective and subjective experiments to show the performance of our proposed framework, and compare against the state-of-the-art offset-cubemap. Our objective experiments show that our quality based approach consistently outperforms the traditional QP based approach under different network conditions. The subjective study shows that $tiled\_cubemap$ offers better or similar perceived quality compared to offset-cubemap while achieving up to 670\% storage savings.

{\tiny
	\bibliographystyle{IEEEtran}
	\bibliography{ref.bib}
}

\end{document}